\documentclass[10pt,a4paper,twoside]{article}
\usepackage{epsfig}
\usepackage{baltlat5}
\usepackage{wrapfig}
\begin{document}
\ \
\vspace{0.5mm}

\setcounter{page}{1}
\vspace{5mm}

\titlehead{Baltic Astronomy, vol.\ts 14, XXX--XXX, 2005.}

\titleb{SEARCHING FOR THE ``MISSING'' PG HOT SUBDWARFS IN SDSS AND GALEX DATA}

\begin{authorl}
\authorb{Richard A.\ Wade}{1}
\authorb{Michele A.\ Stark}{1} 
\authorb{Richard F.\ Green}{2} and
\authorb{Patrick R.\ Durrell}{3}
\end{authorl}

\begin{addressl}
\addressb{1}{Department of Astronomy \& Astrophysics,
Pennsylvania State University, 525 Davey Lab, University Park PA 16802, USA}

\addressb{2}{NOAO, P.O. Box 26732, Tucson AZ 85726-6732, USA}

\addressb{3}{Dept of Physics \& Astronomy, Youngstown State University,
Youngstown OH 44555-2001, USA}
\end{addressl}

\submitb{Received 2005 July ??}

\begin{abstract}
Many EHB stars have been found in short--period binaries, where the
companions in these post--common envelope systems are either white
dwarfs or dM stars; these systems are catalogued as hot subdwarfs
because the sub\-dwarf is the more luminous component.  Hypothesized
Roche--lobe overflow systems (with more massive companions) may
largely be uncatalogued, since the G band or Ca II K--line from the
companion may have caused them to be overlooked or discarded.  In
particular, many candidate objects were excluded from the PG catalog
because of such spectroscopic indicators.  Could these rejects include
large numbers of ``missing'' hot subdwarfs? We have examined 2MASS,
SDSS, and GALEX archival data for large subsets of these rejected
stars, and conclude that only a handful (about 3\%) show indications
of binarity; most are consistent with (single) metal--poor F stars, as
was originally supposed.
\end{abstract}

\vskip4mm


\resthead{\LaTeX\ style for Baltic Astronomy}{R.~A.~Wade, M.~A.~Stark,
R.~F.~Green, P.~R.~Durrell}

\sectionb{1}{HYPOTHESIZED ``MISSING sdB'' STARS}

Binary population synthesis (BPS) models suggest that many
core--helium burning, thin hydrogen envelope objects (``sdB stars'')
are not yet discovered, hidden in binary systems with luminous
(non--degenerate) relatively cool companions. The existing catalogs of
sdB stars would therefore be significantly biased, and a clear picture
of the true situation regarding formation and current population of
sdB stars is lacking.  It is desirable to learn whether these proposed
hidden populations of sdB stars actually exist.

It has been suggested (Han et al.\ 2002, 2003) that the stars that
were rejected from the Palomar--Green (PG: Green, Schmidt \& Liebert
1986, GSL86) survey for ultraviolet--excess (UVX) objects are a
potential rich source for some types of these.  During the PG survey,
candidate UVX objects were those with (transformed) $U-B < -0.46$.
Because of the large errors in $U-B$, $\sigma\approx 0.38$, the color
selection was supplemented by classification spectroscopy, for more
accurate temperature information. Many candidate UVX targets were
indeed excluded from the final PG catalog, because their spectra
showed the Ca II K line (or the G band) in absorption. These rejected
K--line stars were inferred to be metal--poor subdwarf F or G stars
that crept into the candidate list owing to a combination of low
metal--line blanketing and photometric errors.

In the alternative BPS view, these ``PG--rejects'' could be binaries
containing sdB or sdO stars.  The cool star would contribute a K line
and make the blended $U-B$ color marginal for the PG color criterion,
giving a result similar to a metal--poor subdwarf.  In this
interpretation, therefore, the PG--rejects actually belong in the PG
catalog, and moreover would constitute important evidence in favor of
certain binary formation channels for sdBs.

\sectionb{2}{THE SAMPLE OF REJECTED (K--LINE) STARS}

Here we use broad--band photometry over extended wavelength ranges to
assess two large subsamples from the list of PG rejects.  This allows
the spectral energy distribution of a composite hot+cool system to be
distinguished from that of a single star.  From the master catalog of
1125 PG--rejects (RW \& RFG, in preparation), we found 291 stars that are
present in both the Two--Micron All--Sky Survey (2MASS) Point Source
Catalog and the Sloan Digital Sky Survey (SDSS) DR2 survey region, and
we identified 103 stars that have GALEX observations accessible in the
first Data Release. These samples are almost completely
non--overlapping.

\sectionb{3}{THE SDSS/2MASS DATASET: COLOR--COLOR DIAGRAMS}

Of the 291 K--line stars that overlap 2MASS/SDSS, we consider here
the 173 stars with Sloan $r$ magnitudes in the range 14.00 to 16.00
(median $r=14.86$) that have non--flagged (unsaturated) magnitudes in
at least 4 of the 5 passbands of the SDSS ($ugriz$) photometric system
(136 have all 5 optical magnitudes). All of these 173 stars have
detections in the 2MASS $J$, $H$, and $K_s$ bands,

We plotted the 173 K--line stars in both $(g-r, u-g)$ and $(r-K_s,
g-r)$ color--color diagrams, along with 199 PG stars that have been
classified as hot subdwarfs.  We compared these with loci for the Pop
I main sequence, metal--poor main sequence, and metal poor giants
(representing horizontal branch stars).  We also considered three
sequences of composite (binary) models.  These combine the light from
a hot subdwarf star ($T_{\rm eff} = 25000$~K, 30000~K, and 35000~K; $M_V$
derived from the zero--age EHB calculations of Caloi 1972) with the
light from a cool main--sequence (MS) companion.  These sequences
emerge from the hot end of the stellar locus (faintest, coolest
companions at this end), loop away from the single--star locus, and
then loop back to meet the stellar locus at a (single--star) $T_{\rm
eff} \approx 10000~{\rm K}$. At this end, the `cool' (A or F star)
companion dominates.

In these diagrams, the recognized PG hot subdwarfs and the K--line
PG--rejects are very different groups of stars.  In this sense, the
spectroscopy carried out by GSL86 succeeded in improving on the
photographic $U-B$ color selection.  Some catalogued PG hot subdwarfs
clearly are composite objects (Stark \& Wade 2003; Reed \& Stiening
2004).  Most PG--reject stars, however, are consistent with being
single stars, just as they were interpreted to be by GSL86.  Except
for a few outliers, they are not EHB+MS  binary systems.

\sectionb{4}{FITTING THE PG--REJECT STARS AS SINGLE STARS}

We fitted the observed magnitudes for the 173 stars with model
magnitudes, derived from the synthetic photometry done by the Padova
group (Girardi et al.\ 2002, 2004).  We considered all 
Padova models with $T_{\rm eff}$ in the range 4000 -- 50000~K, 
$4.0 < \log g < 5.0$, and metallicities between solar and
[M/H] = $-2.5$.

For each of the 173 stars, we scaled each model in brightness to find
the best fit.  We chose as the best overall model, the one that gave
the smallest reduced chi--square statistic, $\chi^2_\nu$.  Seven
outliers have either large $\chi^2_\nu$ or unusual $T_{\rm eff}$ or
$\log g$. All of the remaining 166 PG--rejects can be fitted as single
stars with $T_{\rm eff}$ in the range 5000 -- 7100~K.  The
$\chi^2_\nu$ values for these 166 non--outliers are acceptably small,
given our present understanding of the SDSS error estimates, our
neglect of interstellar reddening, etc.  Most of these stars (136 of
166) are preferably fitted with low--metallicity models, [M/H] =
$-1.0$ or below, consistent with the GSL86 interpretation that these
are metal--poor F and G subdwarfs.

Two of the outliers have SDSS spectra.  They show Mg~Ib, Na D, and Ca
II infrared triplet absorption, but the continua are blue. Both stars
lie among the composite models in the color--color diagrams. A
plausible model for the first is a 30000--35000~K hot subdwarf plus a
$T_{\rm eff}\approx 6000$~K MS star.  A plausible model for
the second outlier is a $\approx 30000$~K hot subdwarf plus a $T_{\rm
eff}\sim 7500$~K MS star.  The other outliers lie either
close to the hot single--star locus or the sequences of composite
models; one may be a blue horizontal branch star.

\sectionb{5}{THE GALEX DATASET}

 The GALEX photometry is in two bands, Far-- and Near--Ultraviolet
$(F, N)$, with $\lambda_{\rm eff} = 1528$\AA\ and 2271\AA.  Figure 1
is a two--color diagram, with models of single and composite stars
shown (synthetic photometry from Kurucz models).  At the hot (upper
left) end, we show the observed colors of six known EHB stars (three
of these are in binary systems, based on their 2MASS colors).  We also
found GALEX observations for a number of well--observed ($V\sim 9$)
metal--poor single stars near $T_{\rm eff}=6000$~K, and thus
determined that the locus for such stars actually lies higher than the
model line shows, by about 1--2 mag in $F-N$.

Of the 103 PG--rejects that were observed by GALEX, only twelve were
detected in both FUV and NUV bands.  We derived limits on $F-N$ for
the rest (assuming $F > F_{\rm lim} = 19.9$). We plot 24 limits in
Figure 1; the other 67 systems have similar locations in the figure
and are omitted to reduce confusion.  Only three of the 103 stars show
far--UV flux consistent with the presence of a hot star; the rest are
consistent, given the errors and limits, with single cool stars.  All
of the ``hot'' detections have red $J-K_s$ colors from 2MASS, so these
stars are composite.  (One of the three ``hot'' cases, at $N-V \approx
5.1$ and $F-N = 3.0$ is not an EHB+MS binary but can be modeled as a
$T_{\rm eff}\approx 30$~kK WD+dG system.)

\sectionb{6}{SUMMARY}

A few objects ($\approx 3$\%) in our sample of PG--reject stars may
plausibly be binary systems that include a hot subdwarf star as a
member.  The vast majority of the PG--reject stars, however, are
sufficiently modeled as single stars, consistent with their being the
metal--poor sdF and sdG contaminants that GSL86 were guarding
against. The color--color sequences of sdB + cool (MS) star binaries
are well separated from the observed colors of the PG--reject stars in
both the optical--infrared plane and the optical--ultraviolet plane.
There is no compelling evidence for large numbers of additional hot
subdwarf stars hiding in binaries that were rejected from the PG
catalog.

\vskip4mm
\vbox{
\centerline{\psfig{figure=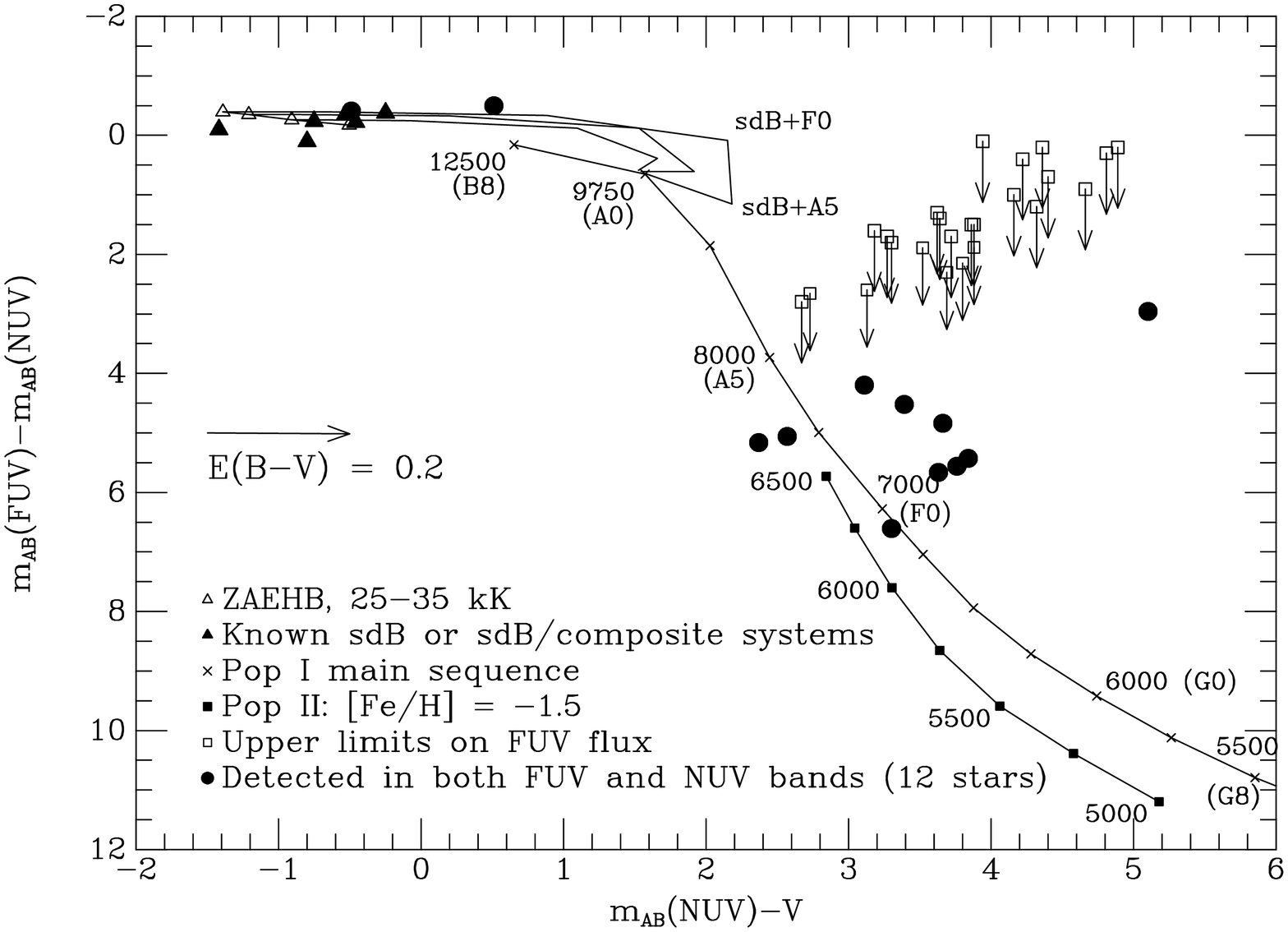,width=84truemm,angle=-90,clip=}}
\captionc{1}{GALEX--Visible color--color diagram.}
}
\vskip4mm

ACKNOWLEDGMENTS. We made use of SDSS Data Release 2, the 2MASS, GALEX
Data Release 1, the USNO-A2 catalog, and SIMBAD.  Supported in part by
Sigma Xi, NASA grants NAG5-9586 and NGT5-50399, and the Pennsylvania
Space Grant Consortium.

\goodbreak

\References

\refb
Caloi, V.\ 1972, A\&A 20, 357

\refb
Girardi, L. et al.\ 2002, A\&A 391, 195

\refb
Girardi, L, Grebel, E.K., Odenkirchen, M.\ \& Chiosi, C.\ 2004, A\&A 422, 205

\refb
Green, R.F., Schmidt, M.\ \& Liebert, J. 1986, ApJS 61, 305 (GSL86)

\refb
Han, Z., Podsiadlowski, Ph., Maxted, P.F.L., Marsh, T.R. \& Ivanova, N. 2002, MNRAS 336, 448

\refb
Han, Z., Podsiadlowski, Ph., Maxted, P.F.L.\ \& Marsh, T.R. 2004, MNRAS 341, 669

\refb
Reed, M.D.\ \& Stiening, R.\  2004, PASP 116, 506

\refb
Stark, M.A.\ \& Wade,  R.A.\  2003, AJ 126, 1455


\end{document}